\documentstyle[preprint,aps]{revtex}
\tightenlines
\begin{document}
\title{Exploratory Calculations of the Effects of Higher Shell Admixtures on 
Static Electric Quadrupole and Magnetic Dipole Moments of Excited States}

\author{S. Aroua$^1$, L. Zamick$^2$, and M.S. Fayache$^1$\\
(1) D\'{e}partement de Physique, Facult\'{e} des Sciences de Tunis\\
Tunis 1060, Tunisia\\
\noindent (2) Department of Physics and Astronomy, Rutgers University
Piscataway, New Jersey 08855}
\date{\today}
\maketitle
\begin{abstract}
Using the interaction $Q \cdot Q ~+~ xV_{S.O.}$ where $V_{S.O.}$ is a two-
body spin-orbit interaction, we study the effects of varying $x$ on the static 
electric quadrupole and magnetic dipole moments of the $2^+_1$ and $2^+_2$ 
states of $^{10}Be$. This is done both in the valence space $s^4p^6$ and in a 
larger space in which 2 $\hbar \omega$ excitations are allowed. In the former 
case, for $x=0$, we have the Wigner Supermultiplet limit in which the 
$2^+_1$ and $2^+_2$ states are degenerate and correspond to $K=0$ and $K=2$ 
rotational states, with equal and opposite static quadrupole moments. Turning 
on the spin-orbit interaction with sufficient strength in the valence space 
gives an energy splitting to the two $2^+$ 
states in accord with experiment. When higher shell admixtures are allowed, we 
get quite a different behaviour as a function of $x$ than in the valence 
space. Of particular interest is a value of $x$ for which $Q(2^+_1)$ and 
$Q(2^+_2)$ both nearly vanish, and so does (somewhat coincidentally) 
$\mu(2^+_1)$.
\end{abstract}

\section{Introduction}

In this work, we make exploratory calculations of the effects of higher 
shell admixtures on nuclear properties. We will focus on the static electric 
quadrupole and magnetic dipole moments of the excited states of even-even 
nuclei. We shall use $^{10}Be$ as a kind of playing field to see if any new 
or interesting phenomena emerge. Of course, $^{10}Be$ is of interest in its 
own right, and experimental measurements of its excited states may become 
feasible in the near future.

Why $^{10}Be$? First of all, it is light enough that we $can$ do shell model 
calculations with higher shell admixtures. Secondly, because this is a nucleus 
for which $N \neq Z$, some non-trivial aspects emerge. For instance, whereas 
in $^8Be$ the magnetic moment of the $2^+$ state is isoscalar and hence less 
sensitive to the details of the nucleon wave function, in $^{10}Be$ one also 
picks up isovector terms which are very sensitive to such details.

In doing the higher shell calculations, we would like to distinguish between 
'expected' effects and surprising effects. An example of the former is the 
fact that quadrupole properties such as $B(E2)$ get enhanced by higher shell 
admixtures. Indeed, the justification for using effective $E2$ charges in a 
valence space comes from perturbative calculations in which $\Delta N=2$ 
admixtures are allowed. But, as we will see in the following sections, there 
are some new and less obvious effects. 

We call this work exploratory in part because we hope that, although we limit 
our calculations to $^{10}Be$, ideas will emerge that will be relevant to 
other nuclei. Also, since we limit our excitations to $\Delta N=2$, we cannot 
be sure what even higher shell admixtures will do.

\section{The Interaction and the Limits}

We shall use the interaction $Q \cdot Q ~+~ xV_{S.O.}$, where 

\[Q \cdot Q \equiv \sum_i (\frac{p^2}{2m} + \frac{1}{2}m \omega^2 r_i^2)
-\chi \sum_{i < j}Q(i) \cdot Q(j) \]

\noindent (we do not include $i=j$ terms), and where $V_{S.O.}$ is a two-body 
spin-orbit interaction that was introduced by Zheng and Zamick \cite{ann}. The 
interactions $Q \cdot Q$ and $V_{S.O.}$ have been previously used in various 
calculations \cite{qq0,qq1}. For $x=1$, Zheng and Zamick got the best fit to 
$G$ matrix elements of realistic interactions, but it has been found that 
several properties in the nuclear medium could be better fit with $x=1.5$
\cite{a14}. 
We do $not$ introduce a one-body spin-orbit term. Rather, the spin-orbit 
splitting is implicitly generated by the two-body spin-orbit interaction. 
Furthermore, we will fix the strength $\chi$ of the 
$Q \cdot Q$ interaction so that, when used in the hamiltonian given above 
with the spin-orbit coupling strength $x$ taken to be equal to its standard  
of $x=1$, the excitation energy of the first $2^+$ state comes out to be  
closest to the experimental value of 3.368 $MeV$. This is achieved by taking 
$\chi = 0.5241~MeV/fm^4$ in the valence space calculations and $\chi = 0.4157
~MeV/fm^4$ in the (0+2) $\hbar \omega$ calculations. 
We will then study the 
effects of varying $x$ on the static electric quadrupole and magnetic dipole 
moments of the $2^+_1$ and $2^+_2$ states. 

\subsection{The small $x$ limit}

For $x=0$ in the valence space, we get the Wigner Supermultiplet limit of the 
$0p$ shell as nicely discussed in the book by Hammermesh \cite{ham} (see also 
\cite{qq0}). In that limit, the first $2^+$ state of $^{10}Be$ is doubly 
degenerate with quantum numbers [420]. We can also use the $SU(3)$ quantum 
numbers $(\lambda, \mu)=(2,2)$. In fact, the double degeneracy comes from 
any Wigner plus Majorana interaction, of which $Q \cdot Q$ is a special case. 

From the point of view of the rotational model, the $2^+_1$ state is a member 
of the $K=0$ ground state band, and the $2^+_2$ state is the lowest member 
of a $K=2$ band. They both have the same $intrinsic$ electric quadrupole 
moment $Q_0$, but they have equal and opposite $laboratory$ electric 
quadrupole moments $Q$. This can be seen from the rotational formula \cite{bm}

\[ Q = \frac{ \left [ 3K^2 -I(I+1) \right ]}{(I+1)(2I+3)} Q_0 \]

\noindent Thus, the small $x$ limit is characterised by $E(2^+_1)=E(2^+_2)$ 
and by $Q(2^+_1)=-Q(2^+_2)$.

Experimentally, the $2^+_1$ and $2^+_2$ states are $not$ degenerate. 
The $2^+_1$ state is at 3.368 $MeV$, and the $2^+_2$ state is at nearly 
twice the excitation energy of 5.960 $MeV$. Undoubtedly, this is due to the 
importance of the spin-orbit interaction, and hence the need to include  
the latter in our hamiltonian, if only to reproduce energy levels that are 
close to experiment.

\subsection{The large $x$ limit}

When the spin-orbit splitting becomes very large, one gets the $jj$ coupling 
limit in the small space. The wave function for $^{10}Be$ is then 

\[(p_{3/2})^2_\pi (p_{3/2})^4_\nu\]

\noindent Note that the neutron shell is closed in this limit and therefore 
the contribution of the neutrons to the quadrupole moment $Q$ and to the 
magnetic moment $\mu$ is zero. The value of $Q$ for the two protons is also 
zero. This is a rather special case where the numbers of holes in the 
$p_{3/2}$ shell equals the number of particles. In general, 

\[Q(j)=-Q(j^{-1})\]

\noindent The only way this can be satisfied for $(p_{3/2})^{J=2}$ is if $Q$ 
is zero. This is a peculiarity of our choice of $^{10}Be$ as the playing 
field. But it gives us a bonus in that we have a very simple limit at large 
$x$ which will guide us as we go from small to large $x$. 

As for the magnetic dipole moment, it is easy to evaluate its value in the 
large $x$ limit. We simply have two $p_{3/2}$ protons coupled to $J=2^+$, and 
using the Schmidt value for a single proton $(p_{3/2})_\pi$.

\[ \mu_{sp}=g_{s,\pi}s+ g_{l, \pi} l\]

\noindent we have for the $(p_{3/2})_{\pi}^2$ $J=2^+$ configuration 

\[\mu(J=2^+)=\frac{J}{j}\mu_{sp}=\frac{2}{3/2}(2.793 + 1)\]

\noindent yielding $\mu(2^+_1) \rightarrow 5.057~\mu_N$ as $x~\rightarrow~
\infty$. This will also be a useful guide in our discussion below, but we 
must remember that all the above limits are for valence space calculations.

\section{Discussions of Tables and Figures}

In Table I, we present the results of electric quadrupole moment calculations 
for various values of $x$, the strength of the spin-orbit interaction. We give 
the results in the small space in Table Ia and in the large space in Table Ib. 
We also show the excitation energies of the two $2^+$ states, as well as the 
values of the electric quadrupole transition amplitudes $B(E2: 0^+_1 
\rightarrow 2^+_1)$ and $B(E2: 0^+_1 \rightarrow 2^+_2)$. Similarly, in 
Tables IIa and IIb, we give the corresponding magnetic dipole moments 
$\mu(spin)$, $\mu(orbital)$ and $\mu(total)$ for the $2^+_1$ and $2^+_2$ 
states. 

We also illustrate the results in various figures as follows. In Fig. 1, we 
give the electric quadrupole moment of the $2^+_1$ state versus $x$ as 
calculated in the small and large spaces. In Fig. 2 we give the small-space 
results for $Q(2^+_1)$ and $Q(2^+_2)$ versus $x$, and in Fig. 3 we give the 
corresponding large-space results. 

In figures 4, 5 and 6 we present a comparison of the magnetic dipole 
moment $\mu(2^+_1)$ as calculated in the small and large model spaces, as a 
function of $x$. In Fig. 4 we have $\mu(2^+_1)_{spin}$, in Fig. 5 
$\mu(2^+_1)_{orbital}$, and in Fig. 6 $\mu(2^+_1)_{total}$. 

\subsection{Discussion of $Q(2^+)$ and $B(E2: 0^+_1 \rightarrow 2^+)$ as a 
function of $x$}

In Fig. 1, we see a large difference in the behaviour of the electric 
quadrupole moment $Q(2^+_1)$ of the $2^+_1$ state versus $x$, especially for 
$x < 1$. In the small space [$(0s)^4(0p)^6$], the behaviour is much simpler. 
$Q(2^+_1)$ is always negative, as we would expect for a prolate nucleus (with 
a positive intrinsic quadrupole moment $Q_0$). We have noted previously that, in the 
$jj$ coupling limit, $Q(2^+_1)$ should vanish, so it is not surprising to 
find that for the most part (except near $x=0$) the value of $Q$ decreases 
steadily from about -4 $e~fm^2$ towards zero as $x$ increases. 

In the large space, on the other hand, $Q(2^+_1)$ starts out large and 
positive ($\approx 5.4~e~fm^2$). It then decreases and goes through zero at 
$x \approx 0.63$, reaches a negative value of about $-5~e~fm^2$ at $x=1.5$, and 
then starts decreasing to zero. Why are the results in the small and large 
space so different? 

Let us first look at the large $x$ behaviour. The fact that the magnitude of 
(negative) $Q(2^+_1)$ is larger in the large space is understandable. The 
effects of including $\Delta N=2$ admixtures is precisely to change bare $E2$ 
charges to effective charges. A popular choice of effective charges is 
$e_p=1.5$ and $e_n=0.5$ for the proton and the neutron, respectively. In the 
calculations, one does not necessarily get exactly these values, but the trend 
is in that direction. Certainly the isoscalar $E2$ gets enhanced. Not only 
$Q(2^+_1)$ but also $B(E2)_{0^+_1 \rightarrow 2^+_1}$ should get enhanced if 
this effective charge argument is correct. From Table I, we see that for 
say $x=1.5$ $B(E2)_{0^+_1 \rightarrow 2^+_1}$ is 18.19 $e^2~fm^4$ in the 
small space, but is equal to 42.63 $e^2~fm^4$ in the large space. This then 
supports the effective charge argument. 

However, at small $x$ the quadrupole moments are of opposite sign in the 
small and large spaces. Recall that in the small space, at $x=0$, the $2^+_1$ 
and $2^+_2$ states are degenerate. We can arrange the two states so that one 
is a $J=2~K=0$ state and the other $J=2~K=2$. The $B(E2)$ to the first state 
should be much larger than to the second. The two states have the same 
intrinsic quadrupole moment in the rotational model, and hence equal and 
opposite quadrupole moments in the laboratory. If we turn on the spin-orbit 
interaction very weakly, the lower energy state is clearly dominantly $K=0$ 
and the upper one $K=2$. For example, for $x=0.1$, the values are: 
$E(2^+_1)=3.352~MeV$, $Q(2^+_1)=-4.042~e~fm^2$ and $B(E2)_{0^+_1 \rightarrow 
2^+_1}=21.34~e^2~fm^4$, whilst for the higher state $E(2^+_2)=3.384~MeV$, 
$Q(2^+_1)=+3.598~e~fm^2$ and $B(E2)_{0^+_1 \rightarrow 
2^+_1}=3.578~e^2~fm^4$. There is a small amount of $K$ mixing due to the spin-
orbit interaction, but not very much. 

In the large space, what appears to be happening right off the bat at $x=0$ 
is that a state which is dominantly $K=2$ is lying lower in energy than one 
which is dominantly $K=0$. 

This is evident from the corresponding values for the (0+2) $\hbar \omega$ 
properties of the $2^+_1$ and $2^+_2$ states obtained with $x=0$:

\noindent $E(2^+_1)=2.485~MeV$, $Q(2^+_1)=5.453~e~fm^2$ and $B(E2)_{0^+_1 
\rightarrow 2^+_1}=5.502~e^2~fm^4$\\
$E(2^+_2)=3.972~MeV$, $Q(2^+_2)=-5.559~e~fm^2$ and $B(E2)_{0^+_1 \rightarrow
2^+_2}=50.48~e^2~fm^4$.

As we turn on the spin-orbit interaction, a more normal behaviour starts to 
be restored, one in which $B(E2)_{0^+_1 \rightarrow 2^+_1}$ is much larger. 
The reason that at $x=0$ the $J=2~K=2$ state lies lower than the $J=2~K=0$ 
is not clear. It must be due to $\Delta N=2$ mixing from the quadrupole-
quadrupole interaction. Such mixing does not maintain the $SU(3)$ symmetry 
of the Elliott hamiltonian (although in first order perturbation theory 
all that happens is that the strength of the $Q \cdot Q$ interaction 
gets renormalized, so that in that approximation one would still get 
Elliott's formula for the energies, and in particular the $2^+_1$ and $2^+_2$ 
states of $^{10}Be$ could still be degenerate. Here, however, we are $not$ 
doing first order perturbation theory, but rather exact (0+2) $\hbar \omega$ 
matrix diagonalization).

\subsection{Discussion of the magnetic dipole moment $\mu(2^+)$ as a function 
of $x$}

There are some surprises for the static magnetic dipole moments as well. As seen 
in Fig. 6, whereas in the small space the magnetic moment of the $2^+_1$ state 
increases steadily with $x$ (and as mentioned previously approaches 5.057 
$\mu_N$ as $x \rightarrow \infty$), in the large space $\mu(2^+_1)$ starts out 
small and positive but decreases to almost zero at $x=0.63$, where there is a 
near cancellation between the spin and orbital parts of the magnetic moment. 

The negative values of $\mu(2^+_1)_{spin}$ from $x=0$ to $x=1.5$ obviously 
come from the neutrons. At $x=0$, $\mu_{spin}$ must be zero because in the 
Wigner Supermultiplet theory the $2^+_1$ state has $S=0$. As $x$ goes to 
infinity, we approach the $jj$ coupling limit, and the neutrons will not 
contribute to $\mu_{spin}$ because they form 
an $S=0$ closed neutron $p_{3/2}$ subshell. Only the protons will contribute, 
in that limit, and $\mu_{spin}$ for $p_{3/2}$ protons is positive. Hence, only 
for some intermediate values of $x$ will the neutrons' contribution to 
$\mu_{spin}$ be able to dominate. 

For large $x$, the results are somewhat easier to understand. The fact that 
$\mu_{spin}(large~space)~<~\mu_{spin}(small~space)$ can be understood by the 
fact that the higher shell admixtures renormalize and considerably enhance 
the effective strength of the $Q \cdot Q$ interaction in the valence space. 
This makes the ratio $x/ \chi$ of the spin-orbit strength to the $Q \cdot Q$ 
strength effectively smaller. Thus, roughly speaking, the value of 
$\mu_{spin}(large~space)$ for a given $x$ in Fig. 4 should be compared with 
a value of $\mu_{spin}(small~space)$ for a smaller $x$ (say $x/1.5$). 
The values of $\mu_{spin}$ will then be much closer. Alternatively, one could 
use a larger value of $\chi$ in the small space calculation in order to 
make the comparison with the large space calculation for the same $x$.

The value of $\mu_{orbital}$ in the small space is remarkably constant as a 
function of $x$, with a value of about $1.4~\mu_N$. However, in the large 
space and at $x=0$ this quantity is much smaller ($\approx ~0.3~\mu_N$). As 
$x$ increases, $\mu_{orbital}$ also increases, and for large $x$, the values 
in small and large spaces are almost the same. 

The spin-orbit interaction is known to give $K$ mixing and therefore to 
bring many complications to analyses which start from the point of view of 
the rotational model. Ironically, here, the spin-orbit interaction with 
sufficient strength simplifies things. It brings down in energy the expected 
dominantly $K=0~J=2^+$ prolate state, the signatures of which are a strong 
$B(E2)$ transition from the ground state and a large negative electric 
quadrupole moment.

Summarizing this section, we can say that when the coupling of the spin-orbit 
interaction is sufficiently strong, things tend to settle down.

\subsection{The critical value $x \approx 0.63$ where $Q(2^+_1)$, $Q(2^+_2)$ 
and $\mu (2^+_1)$ nearly all vanish}

As we increase $x$ from zero to 2 in the small space, nothing unusual 
happens to $Q(2^+_1)$. It has a large negative value for $x=0$, consistent 
with the $SU(3)$ limit and with a prolate nucleus. The value of $Q(2^+_1)$ 
then gradually approaches zero as $x$ approaches infinity, as expected from 
the discussion of the $jj$ coupling limit in the previous section. 

In the large space, the behaviour is drastically different. Now $Q$ starts 
out {\em positive}, goes through zero at about $x=0.63$ and then becomes 
negative. For large $x$, the behaviour is less surprising. The curves for 
both small and large spaces seem to be going to zero. The fact that 
$|Q(2^+_1)|_{large}~>~|Q(2^+_1)|_{small}$ for large $x$ is consistent with 
the fact that core polarization enhances $Q$ and $B(E2)$ $i.e.$ it gives a 
microscopic justification for using effective charges. 

Getting back to the critical region of $x \approx 0.63$, we find that not only 
$Q(2^+_1)$ but also $Q(2^+_2)$ becomes very small in magnitude, and likewise 
$\mu(2^+_1)$. We can perhaps explain why both $Q$'s become small at the same 
time by noting that in the $SU(3)$ limit, the two states $2^+_1$ and $2^+_2$ 
are degenerate and have equal and opposite quadrupole moments. In the 
rotational model, both the $K=0$ and $K=2$ $J=2^+$ states have the same 
{\em intrinsic} quadrupole moment $Q_0$, but the geometric factors cause the 
laboratory $Q$'s to be equal and opposite. 

Let us suppose that the spin-orbit interaction mixes the $K=0$ and $K=2$ 
states:

\[ |2^+_1 \rangle = \alpha |K=0 \rangle ~+~ \beta |K=2 \rangle \]

\[ |2^+_2 \rangle = -\beta |K=0 \rangle ~+~ \alpha |K=2 \rangle \]

\noindent then 

\[ Q(2^+_1) = {\alpha}^2 Q_{K=0} + {\beta}^2 Q_{K=2} + 2\alpha \beta 
\langle K=0| Q | K=2 \rangle \]

\[ Q(2^+_2) = {\beta}^2 Q_{K=0} + {\alpha}^2 Q_{K=2} - 2\alpha \beta 
\langle K=0| Q | K=2 \rangle \]

\noindent Suppose now that $\langle K=0| Q | K=2 \rangle$ is negligible and 
that $Q_{K=0}=-Q_{K=2}$, we then have 

\[ Q(2^+_1) = ({\alpha}^2 - {\beta}^2) Q_{K=0}\] 

\noindent and 

\[ Q(2^+_2) = ({\beta}^2 - {\alpha}^2) Q_{K=0}\] 

\noindent and if we have equal mixing $i.e.$ $\alpha =\beta = 
\frac{1}{\sqrt{2}}$ we get 

\[ Q(2^+_1) = Q(2^+_2) =0\]

To support this $K$-mixing argument, we should examine the values of 
the electric quadrupole transitions $B(E2)_{0^+_1 \rightarrow 2^+_1}$ and 
$B(E2)_{0^+_1 \rightarrow 2^+_2}$. If there were $no$ $K$-mixing, the 
$B(E2)$ to the $K=0~J=2^+$ state should be much stronger than to the 
$K=2~J=2^+$ state. If there $is$ strong $K$-mixing, the $B(E2)$'s to both 
states should be comparable. 

Before looking at the critical region ($x \approx 0.63$), let us look at the 
large space results for a value of $x$ that gives the best energies for the 
$2^+_1$ and $2^+_2$ states. This corresponds to $x=1.5$ where 
$E(2^+_1)=3.513~MeV$ and $E(2^+_2)=6.307~MeV$, and the values of $Q(2^+_1)$ 
and $Q(2^+_2)$ are -5.039 $e~fm^2$ and 5.384 $e~fm^2$, respectively. This 
is consistent with the lowest state being dominantly $K=0$ and the upper one 
$K=2$, and is probably very close to the $experimental$ situation, not only 
for $^{10}Be$ but for most strongly deformed nuclei. Support to this is given 
by looking at the calculated $B(E2)$'s. We find that $B(E2)_{0^+_1 \rightarrow 2^+_1}
= 42.63~e^2~fm^4$ and $B(E2)_{0^+_1 \rightarrow 2^+_2}=0.989~e^2~fm^4$. 

Below the critical $x$, say at $x=0.1$, we have the reverse situation, with 
$Q(2^+_1)$ positive, $Q(2^+_2)$ negative (5.345 and -5.230 $e~fm^2$ 
respectively), and $B(E2)_{0^+_1 \rightarrow 2^+_2}=42.76~e^2~fm^4$ while 
$B(E2)_{0^+_1 \rightarrow 2^+_1}=5.83~e^2~fm^4$.

Clearly then, there is a cross-over around $x \approx 0.63$ or so. As 
mentioned above, the values of $Q(2^+_1)$ and $Q(2^+_2)$ nearly vanish 
in the critical region $x \approx 0.63$ (they are -0.034 and 0.278 
$e~fm^2$ respectively). What about the $B(E2)$'s? The values of these 
are 24.39 $e^2~fm^4$ to the $2^+_1$ state and 23.07 $e^2~fm^4$ to the 
$2^+_2$ state, nearly equal indeed. This lends a clear support to the 
strong $K$-mixing argument.

The fact that $\mu(2^+_1)$ also nearly vanishes where $Q(2^+_1)$ and 
$Q(2^+_2)$ do is probably coincidental. For the magnetic dipole moment we 
find that the spin and orbital contributions nearly cancel in the vicinity 
of $x=0.63$. As mentioned before, $\mu_{spin}$ vanishes at $x=0$. For small, 
finite $x$, the neutrons tend to dominate to give a negative $\mu_{spin}$. At 
very large $x$ however, and at least in the small space, the neutrons again do 
not contribute because they form a closed $p_{3/2}$ subshell. 

\section{Results that are of general interest}

Some of the results that we have obtained are expected, but some are at least 
at first thought surprising. 

The Wigner Supermultiplet theory, and as a special case $SU(3)$, predicts results that 
are in disagreement with experiment both in the valence space and the valence 
plus 2 $\hbar \omega$ space. In the small space, the $2^+_1$ and $2^+_2$ 
states ($K=0$ and $K=2$) are predicted to be degenerate, whereas 
experimentally the $2^+_2$ state is at about $twice$ the excitation energy of 
the $2^+_1$ state (3.368 $MeV$ and 5.960 $MeV$, respectively). Turning on a 
sufficiently strong spin-orbit interaction ($x \approx 1$ to 1.5) corrects 
this situation, and furthermore the lower state is dominantly $K=0$ with a 
strong $B(E2)$ from the ground state, and the upper one is dominantly $K=2$ 
with a weak $B(E2)$ from the ground state. 

In the large space, the situation is even worse in the absence of a spin-orbit 
interaction. The lowest state is dominantly $K=2$ and the upper one is $K=0$. 
As we turn on the spin-orbit interaction, one first gets a crossover in the 
properties of the lowest $2^+$ state: at about $x=0.63$ one gets about equal 
mixing  of $K=0$ and $K=2$ with the result that the electric quadrupole 
moments $Q(2^+_1)$ and $Q(2^+_2)$ both nearly vanish. In other regions of $x$ 
they are both large in magnitude and nearly equal but opposite in sign. 

As we increase $x$ further, we slowly converge to the experimental situation 
(at $x \approx 1.25$) where the $2^+_1$ and $2^+_2$ states are split by about 
3 $MeV$ and the $B(E2)$ to the $2^+_1$ state is much stronger than to the 
$2^+_2$ state. 

Looking at Fig. 1 we see that for the realistic value near $x=1.5$ we are 
safely away from the crossover region  near $x=0.63$. But this example suggests 
that in other nuclei and at other excitation energies, there may be two states 
of the same angular momentum but with different $K$ or other quantum numbers. 
In that case, it appears that one can get unexpected and different effects 
in a valence space calculation as compared with one in which 2 $\hbar \omega$ 
excitations are allowed. For example, as noted here in the small space 
calculation with $x=0.5$, we have $Q(2^+_1)=-3.878~e~fm^2$ and $Q(2^+_2)=
3.662~e~fm^2$, whereas in the large space the values are $Q(2^+_1)=2.109~e~
fm^2$ and $Q(2^+_2)=-1.816~e~fm^2$. These are drastically different values 
and the difference cannot be considered to be the result of perturbative 
corrections. Likewise, for the magnetic dipole moments in the valence space 
we have at $x=0.5$ $\mu(2^+_1)_{spin}=0.017~\mu_N$, $\mu(2^+_1)_{orbit}=
1.295~\mu_N$ and $\mu(2^+_1)_{total}=1.312~\mu_N$, whereas in the large space 
the corresponding values are -0.387, 0.462 and 0.0753 $\mu_N$. 

When we go to larger $x$, things are not completely chaotic. We expect that 
2 $\hbar \omega$ admixtures will cause electric quadrupole moments and 
$B(E2)$'s to be enhanced, according to the effective charge argument that was 
given previously. Let us arbitrarily define an electric quadrupole moment 
enhancement factor as $Q(2^+_1)_{large}$/$Q(2^+_1)_{small}$. The values of 
this enhancement factor for $x=1,~1.25,~1.5,~1.75$ and 2.0 are respectively 
1.41, 1.99, 2.45, 2.76 and 2.93. We should remember that in the small space 
$Q(2^+_1)$ approaches 0 as $x$ becomes very large ($jj$ limit) and this is 
the reason the enhancement factor, as defined above, increases with $x$. 

Let us now consider the corresponding $B(E2)$ enhancement factor 
$B(E2: 0^+ \rightarrow 2^+_1)_{large}$/$B(E2: 0^+ \rightarrow 2^+_1)_{small}$. 
The values of this factor for the same values of $x$ as above are respectively 
2.05, 2.27, 2.34, 2.28 and 2.15. One might at first expect the $B(E2)$ 
enhancement factors to be the squares of (and hence much larger than) the 
quadrupole enhancement factors, but in fact they are smaller. 

In summary, we have found some surprising behaviour which, given the large 
number of nuclei in the periodic table and the large number of excited states, 
may actually be realized. For $^{10}Be$, we feel that we are realistically in 
a region of $x$ where, although the effects of higher shell admixtures are 
very substantial, leading to large $E2$ enhancements and large magnetic dipole 
suppressions, the results are under control. But it may be possible that for 
other nuclei we have, for example, nearly degenerate $K=0$ and $K=2$ states 
such that the higher shell admixtures lead to very surprising results. The 
examples given in this work are the fact that $Q(2^+_1)$ and $Q(2^+_2)$ nearly 
vanish at $x \approx 0.63$ in the large space, whereas in the small space they 
are large and nearly equal and opposite. The near vanishing of $\mu(2^+_1)_
{total}$ at $x \approx 0.63$ in the large space, and the fact that it takes a 
long 
time before $\mu(2^+_1)_{orbital}$ achieves its full value, are two other 
examples. 

\section{Acknowledgements}

We thank N. Koller and H. A. Speidel for demonstrating to us the value of 
studying static moments of excited states.
Support from the U.S. Department of Energy (contract \# DE-FG 02-95ER-40940) 
is greatfully acknowledged by L. Zamick.

\pagebreak
{\large{\bf {Figure Captions\\}}}

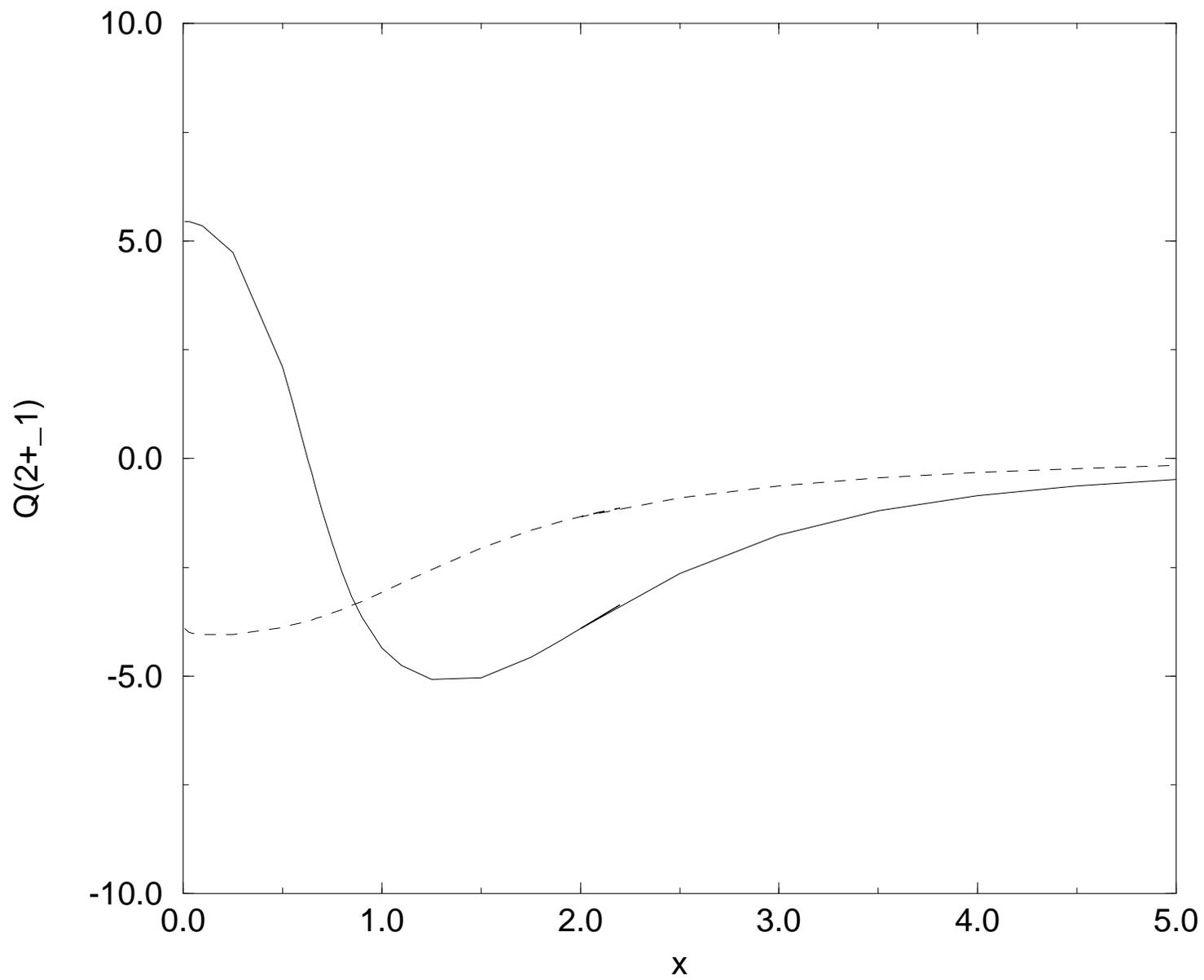
{\bf {Fig. 1:}} The electric quadrupole moment $Q$ of the $2^+_1$ state 
versus $x$ as calculated in the small space (dashed curve) and in the large 
space (solid curve).\\

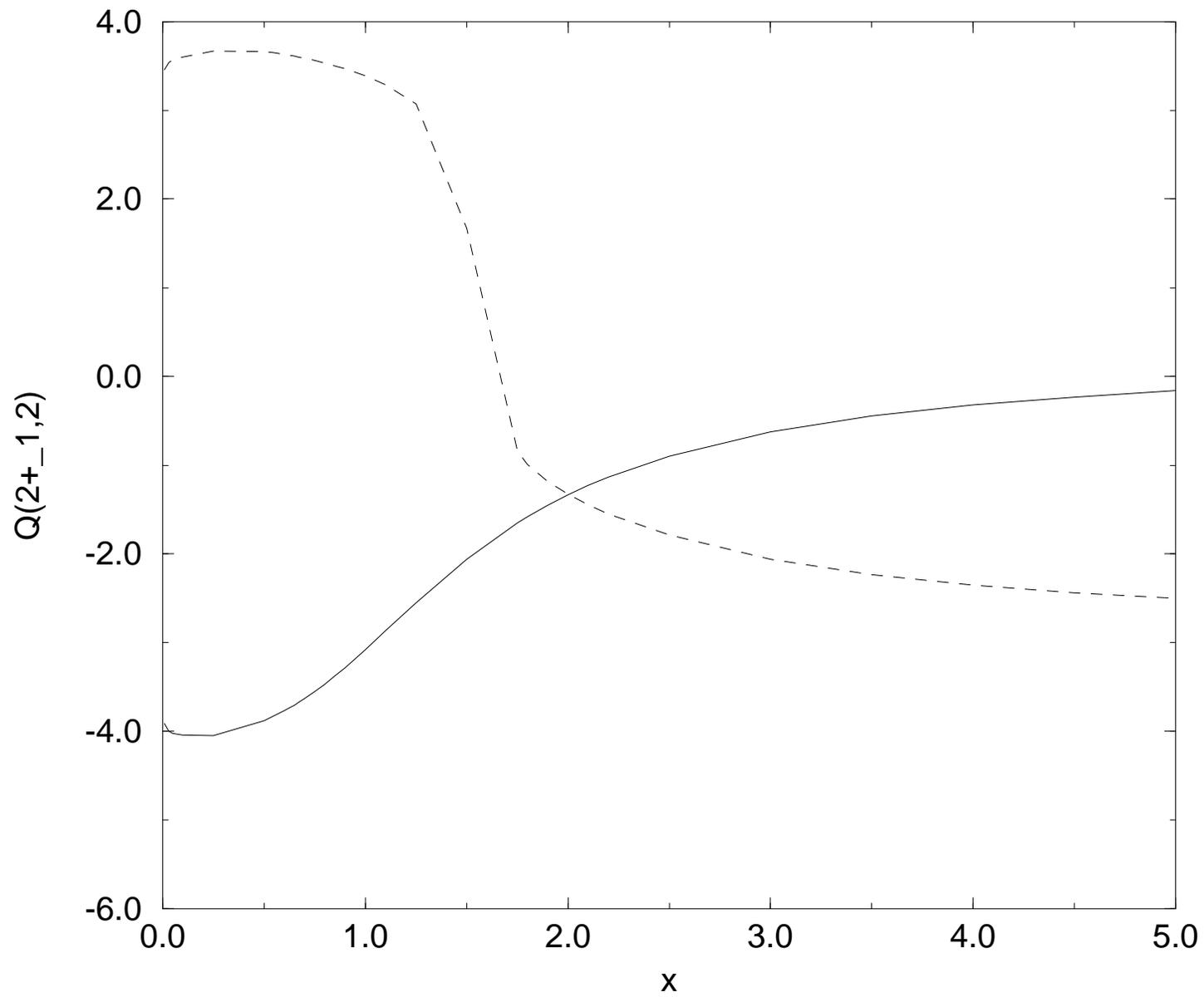
{\bf {Fig. 2:}} Small-space calculations of the electric quadrupole moments of 
the $2^+_1$ state (solid curve) and of the $2^+_2$ state (dashed curve).\\

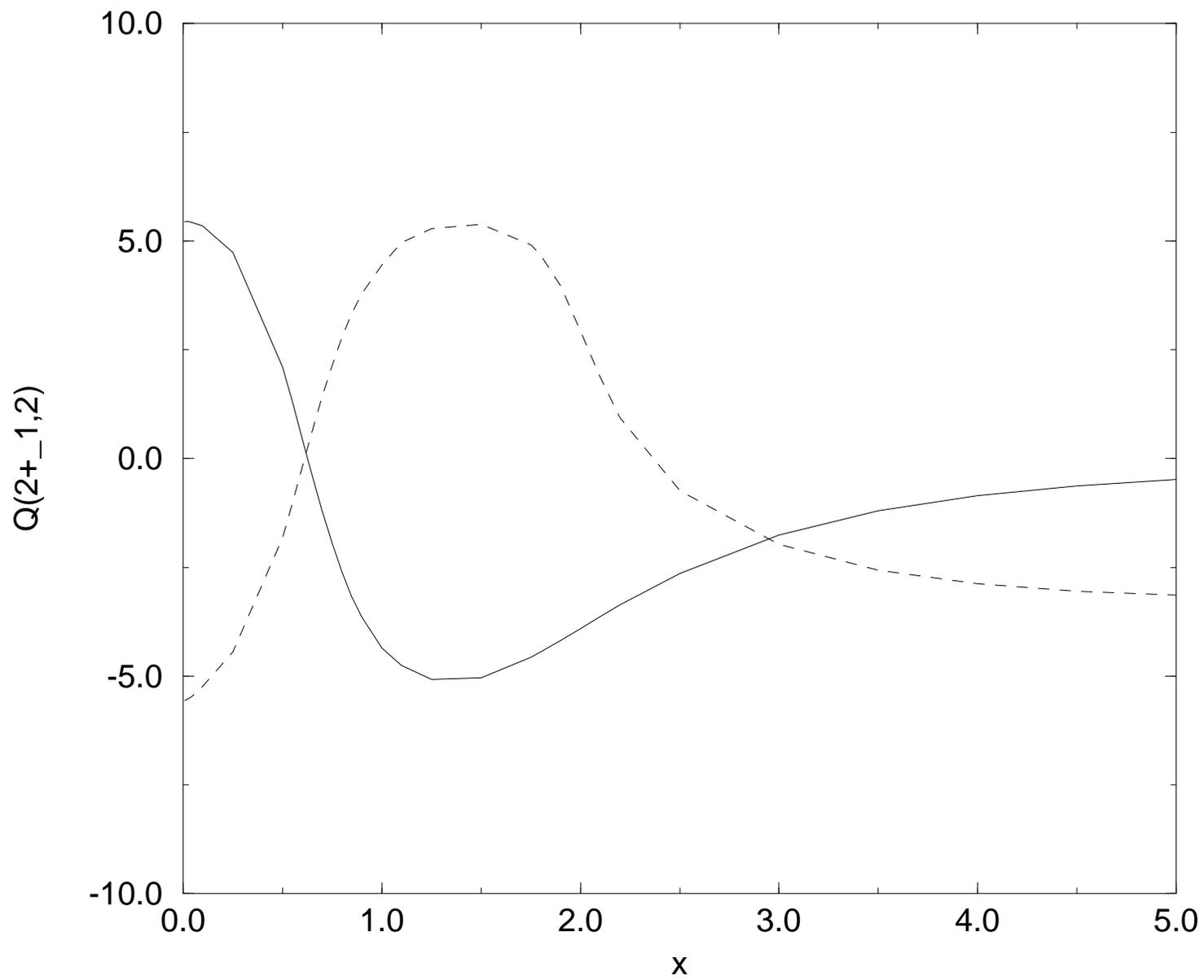
{\bf {Fig. 3:}} Large-space calculations of the electric quadrupole moments of 
the $2^+_1$ state (solid curve) and of the $2^+_2$ state (dashed curve).\\

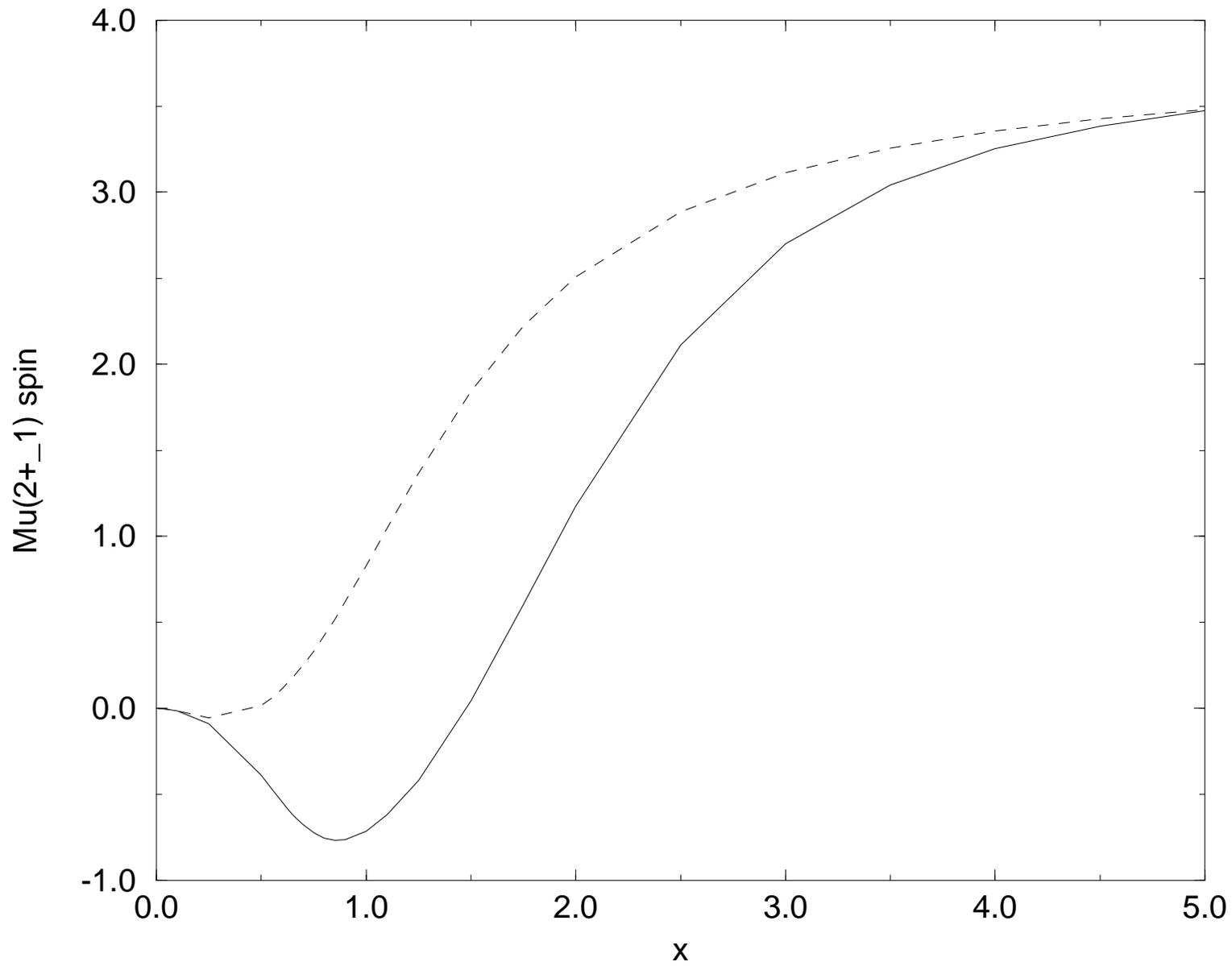
{\bf {Fig. 4:}} The spin component of the magnetic dipole moment $\mu$ of the 
$2^+_1$ state versus $x$, as calculated in the small space (dashed curve) and 
in the large space (solid curve).\\

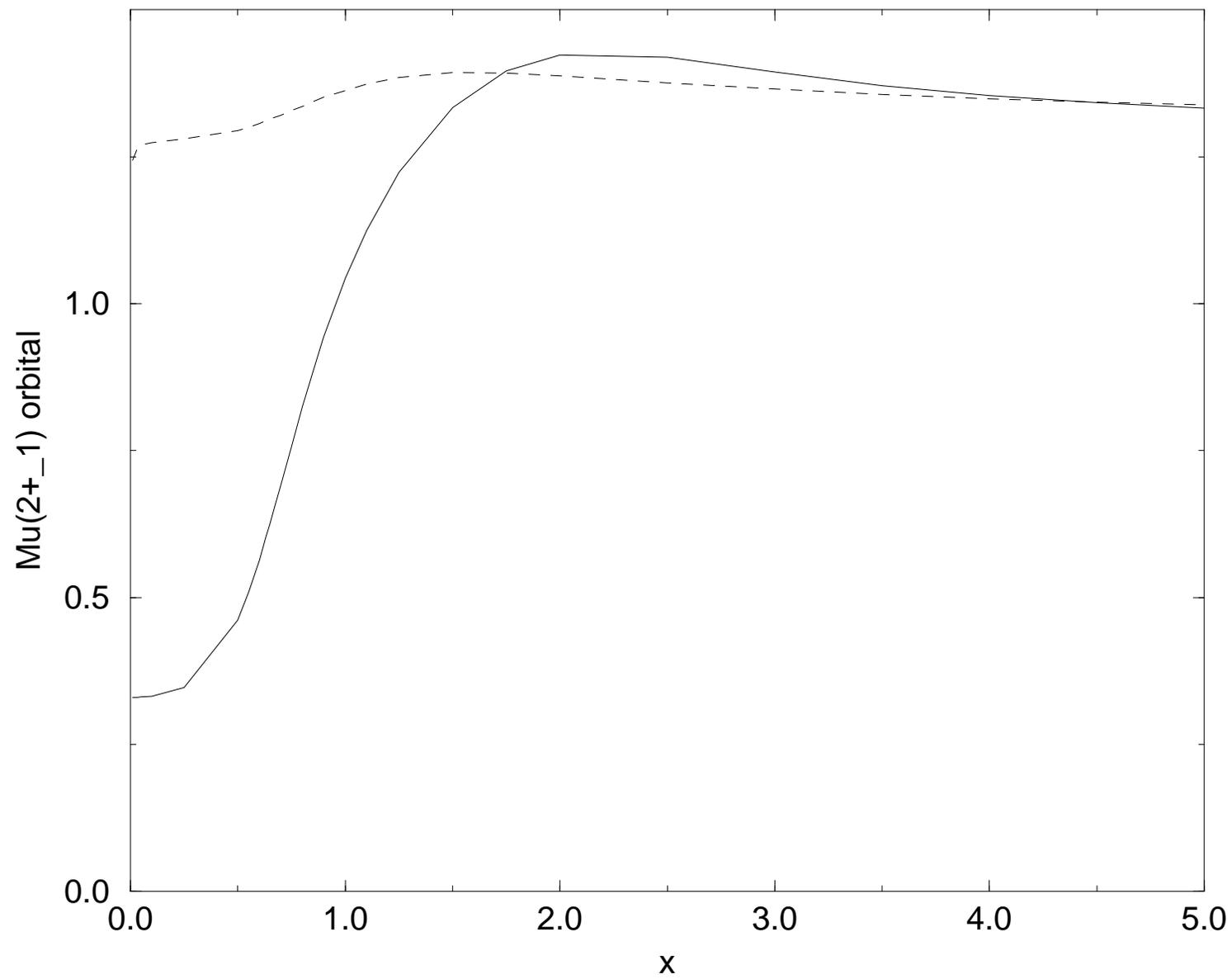
{\bf {Fig. 5:}} Same as Fig. 4, but for the $orbital$ component of the 
magnetic dipole moment $\mu(2^+_1)$.\\

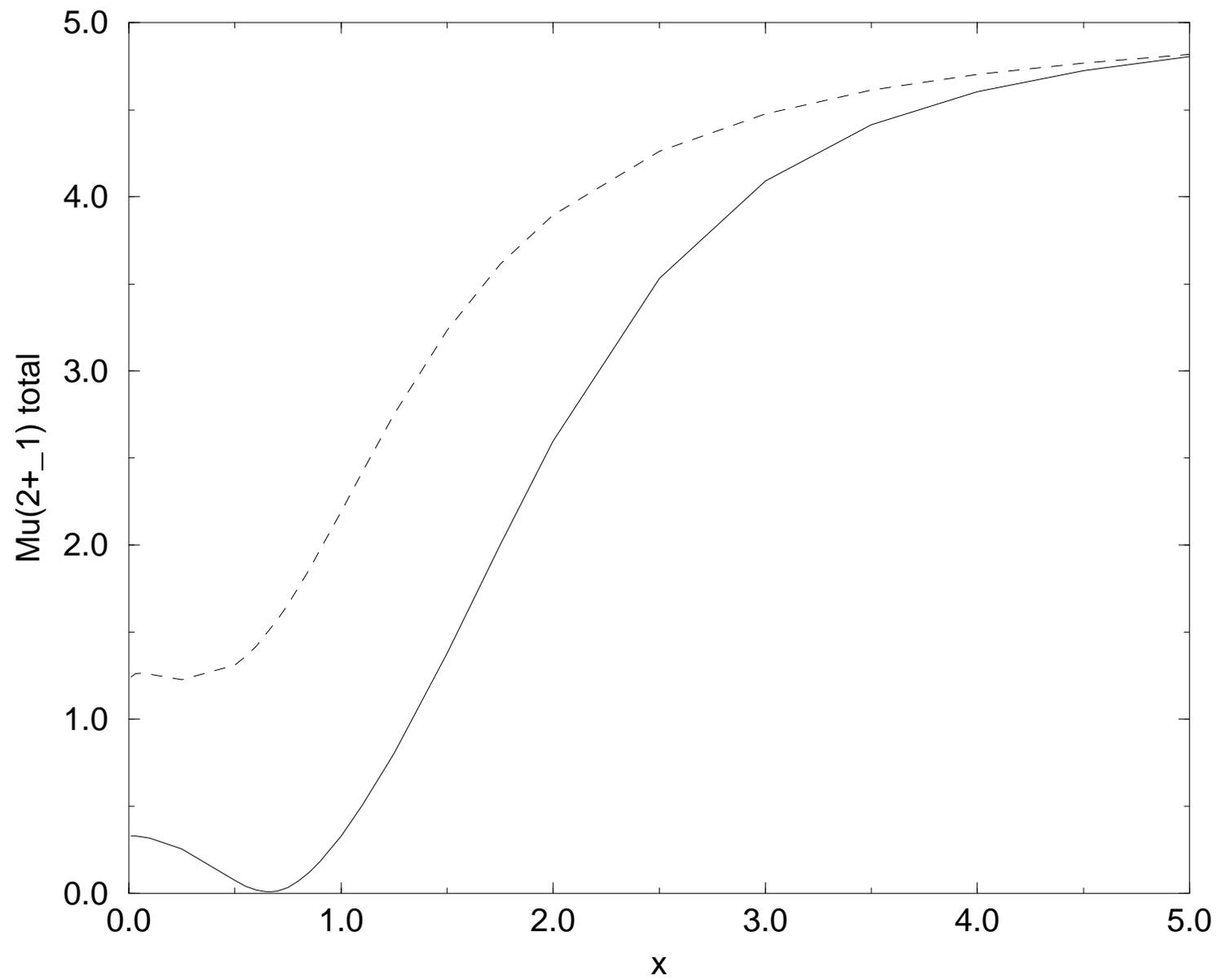
{\bf {Fig. 6:}} Same as Fig. 4, but for the $total$ magnetic 
dipole moment $\mu(2^+_1)$.

\begin{table}

\caption{Excitation energies (in $MeV$), electric quadrupole moments 
$Q$ and transition amplitudes from the ground state $B(E2)$ (in units of 
$efm^2$ and $e^2fm^4$, resp. and using $e_p=1.0$ and $e_n=0.0$), for the 
first two $2^+$ states in $^{10}Be$ as a function 
of the strength $x$ of the spin-orbit term in the interaction $Q \cdot Q + x 
V_{S.O}$, in the small and large shell model spaces.}
\begin{tabular}{cccc|ccc}
   & \multicolumn{3}{c}{(a) 0 $\hbar \omega$ Space} 
& \multicolumn{3}{c}{(b) (0+2) $\hbar \omega$ Space}\\
\tableline
 $x$ & $E(2^+)$ & $Q(2^+)$ & $B(E2: 0^+_1 \rightarrow 2^+)$
& $E(2^+)$ & $Q(2^+)$ & $B(E2: 0^+_1 \rightarrow 2^+)$\\
\tableline
0    & 3.356 & -3.913 & 20.95 & 2.485 & 5.453 & 5.502\\
     & 3.356 &  3.455 & 4.099 & 3.972 & -5.559 & 50.48\\
     &       &        &       &       &        &      \\
0.1  & 3.352 & -4.042 & 21.34 & 2.504 & 5.345 & 5.829\\
     & 3.384 &  3.598 & 3.578 & 3.914 & -5.230 & 42.76\\
     &       &        &       &       &        &      \\
0.25 & 3.338 & -4.046 & 21.24 & 2.608 & 4.744 & 7.738\\
     & 3.532 &  3.668 & 3.137 & 3.819 & -4.445 & 37.30\\
     &       &        &       &       &        &      \\
0.5  & 3.339 & -3.878 & 20.86 & 2.902 & 2.109 & 16.71\\
     & 4.022 &  3.662 & 2.571 & 3.876 & -1.816 & 29.87\\
     &       &        &       &       &        &      \\
0.60 & 3.353 & -3.767 & 20.72 & 3.027 &  0.477 & 22.54\\
     & 4.284 &  3.631 & 2.391 & 3.984 &  -0.217 & 24.70 \\
     &       &        &       &       &        &      \\
0.63 & 3.358 & -3.729 & 20.08 & 3.063 & -0.034 & 24.39\\
     & 4.368 &  3.619 & 2.342 & 4.025 &  0.278 & 23.07 \\
     &       &        &       &       &        &      \\
0.70 & 3.370 & -3.630 & 20.57 & 3.139 & -1.188 & 28.64\\
     & 4.573 &  3.588 & 2.238 & 4.138 & 1.408 & 19.15\\
     &       &        &       &       &        &      \\
0.75 & 3.377 & -3.552 & 20.48 & 3.207 & -1.939 & 31.44\\
     & 4.725 &  3.562 & 2.171 & 4.232 &  2.141 & 16.51\\
     &       &        &       &       &        &      \\
1.0  & 3.403 & -3.080 & 19.88 & 3.379 & -4.348 & 40.68\\
     & 5.552 &  3.389 & 1.952 & 4.825 & 4.456 & 7.391\\
     &       &        &       &       &        &      \\
1.25 & 3.382 & -2.549 & 19.05 & 3.486 & -5.078 & 43.32\\
     & 6.441 & 3.072 & 2.041 & 5.540 &  5.282 & 3.067 \\
     &       &        &       &       &        &      \\
1.50 & 3.312 & -2.057 & 18.19 & 3.513 & -5.039 & 42.63 \\
     & 7.331 & 1.667 & 3.427 & 6.307 & 5.384  & 0.989 \\
     &       &        &       &       &        &      \\
1.75 & 3.203 & -1.651 & 17.42 & 3.456 & -4.565 & 39.78 \\
     & 8.043 & -0.843 & 4.853 & 7.073 & 4.908  & 0.010 \\
     &       &        &       &       &        &      \\
2.0  & 3.070 & -1.334 & 16.78 & 3.316 & -3.910 & 36.00 \\
     & 8.691 & -1.328 & 5.153 & 7.727 & 2.925 & 1.860 \\
     &       &        &       &       &        &      \\
3.0  & 2.459 & -0.626 & 15.31 & 2.266 & -1.752 & 23.06 \\
     & 11.519 & -2.060 & 6.162 & 9.506 & -1.967 & 14.15 \\
     &       &        &       &       &        &      \\
4.0  & 1.843 & -0.322 & 14.67 & 1.097 & -0.856 & 17.06\\
     & 14.563 & -2.352 & 6.551 & 11.870 & -2.878 & 15.30\\
\tableline
\end{tabular}
\end{table}

\begin{table}
\caption{Same as in Table I but for the magnetic dipole moments of the 
first two $2^+$ states in $^{10}Be$ as a function 
of the strength $x$ of the spin-orbit term in the interaction $Q \cdot Q + x 
V_{S.O}$, in the small and large shell model spaces.}
\begin{tabular}{ccccc|cccc}
   & \multicolumn{4}{c}{ (a) 0 $\hbar \omega$ Space} 
& \multicolumn{4}{c}{ (b) (0+2) $\hbar \omega$ Space}\\
\tableline
 $x$ & $E(2^+)$ & $\mu_{spin}(2^+)$ & $\mu_{orbital}(2^+)$ & 
$\mu_{total}(2^+)$ & $E(2^+)$ & $\mu_{spin}(2^+)$ & $\mu_{orbital}(2^+)$ & 
$\mu_{total}(2^+)$ \\
\tableline
0    & 3.356 & 0 & 1.244 & 1.244  & 2.485 & 0 & 0.331 & 0.331 \\
     & 3.356 & 0 & 0.756 & 0.756  & 3.972 & 0 & 1.557 & 1.557 \\
     &       &        &       &       &        &      &      &  \\
0.1  & 3.352 & -0.014 & 1.274 & 1.259 & 2.504 & -0.014 & 0.332 & 0.318 \\
     & 3.384 & 0.005 & 0.719 & 0.724  & 3.914 & -0.602 & 1.412 & 0.810 \\
     &       &        &       &       &        &      &      &  \\
0.25 & 3.338 & -0.053 & 1.281 & 1.228 & 2.608 & -0.090 & 0.347 & 0.257 \\
     & 3.532 & 0.029 & 0.686  & 0.715 & 3.819 & -0.968 & 1.308 & 0.340 \\
     &       &        &       &       &        &      &      &  \\
0.5  & 3.339 & 0.017 & 1.295 & 1.312 & 2.902 & -0.387 & 0.462 & 0.075 \\
     & 4.022 & 0.106 & 0.636 & 0.742 & 3.876 & -0.707 & 1.205 & 0.498 \\
     &       &        &       &       &        &      &      &  \\
0.60 & 3.353 & 0.116 & 1.307 & 1.422 & 3.027 & -0.544 & 0.564 & 0.020 \\
     & 4.284 & 0.144 & 0.618 & 0.762 & 3.984 & -0.499 & 1.120 & 0.621 \\
     &       &        &       &       &        &      &      &  \\
0.63 & 3.358 & 0.153 & 1.311  & 1.464  & 3.063 & -0.588 & 0.601 & 0.013 \\
     & 4.368 & 0.156 & 0.613  & 0.769  & 4.025 & -0.438 & 1.089 & 0.652  \\
     &       &        &       &       &        &      &      &  \\
0.70 & 3.370 & 0.253 & 1.321  & 1.574 & 3.139 & -0.677 & 0.692 & 0.015 \\
     & 4.573 & 0.186 & 0.601  & 0.787 & 4.138 & -0.297 & 1.011 & 0.713 \\
     &       &        &       &       &        &      &      &  \\
0.75 & 3.377 & 0.335 & 1.329 & 1.663 & 3.207 & -0.723 & 0.759 & 0.036 \\
     & 4.725 & 0.207 & 0.593 & 0.800 & 4.232 & -0.210 & 0.954  & 0.744 \\
     &       &        &       &       &        &      &      &  \\
1.0  & 3.403 & 0.830 & 1.363 & 2.193  & 3.379 & -0.714 & 1.044 & 0.331 \\
     & 5.552 & 0.331 & 0.566 & 0.897  & 4.825 & 0.041 & 0.717 & 0.759 \\
     &       &        &       &       &        &      &      &  \\
1.25 & 3.382 & 1.368 & 1.385 & 2.752  & 3.486 & -0.416 & 1.224 & 0.807 \\
     & 6.441 & 0.564 & 0.583 & 1.147  & 5.540 & 0.101 & 0.562 & 0.662\\
     &       &        &       &       &        &      &      &  \\
1.50 & 3.312 & 1.844 & 1.393 & 3.236  & 3.513 & 0.045 & 1.334 & 1.379 \\
     & 7.331 & 1.597 & 0.835 & 2.432  & 6.307 & 0.106 & 0.455 & 0.561  \\
     &       &        &       &       &        &      &      &  \\
1.75 & 3.203 & 2.221 & 1.392 & 3.613  & 3.456 & 0.604 & 1.396 & 2.001 \\
     & 8.043 & 2.519 & 1.091 & 3.611  & 7.073 & 0.207 & 0.374 & 0.581 \\
     &       &        &       &       &        &      &      &  \\
2.0  & 3.070 & 2.507 & 1.388 & 3.895  & 3.316 & 1.176 & 1.423 & 2.599 \\
     & 8.691 & 2.209 & 1.091 & 3.300  & 7.727 & 0.880 & 0.420 & 1.300 \\
     &       &        &       &       &        &      &      &  \\
3.0  & 2.459 & 3.112 & 1.365 & 4.477  & 2.266 & 2.699 & 1.394 & 4.093\\
     & 11.519 & 1.620 & 1.159  & 2.779 & 9.506 & 1.488 & 0.952 & 2.440  \\
     &       &        &       &       &        &      &      &  \\
4.0  & 1.843 & 3.356 & 1.349 & 4.704  & 1.097 & 3.252 & 1.354 & 4.605 \\
     & 14.563 & 1.450 & 1.210 & 2.660 & 11.870 & 1.254 & 1.113 & 2.367 \\
\tableline
\end{tabular}
\end{table}

\end{document}